\documentclass[aps,epsf,preprint]{revtex4}
\usepackage{bm}
\begin{document}
\newcommand{\nv}{{\bf n}}
\newcommand{\xv}{{\bf x}}
\newcommand{\Mv}{{\bf M}}
\newcommand{\Av}{{\bf A}}
\newcommand{\av}{{\bf a}}
\newcommand{\sigmav}{{\bm \sigma}}
\newcommand{\ra}{{\rightarrow}}
\newcommand{\tr}{{\rm tr}}
\title{
Permanent current from non-commutative spin algebra  } 
\author{Gen Tatara}
\author{Hiroshi Kohno} 
\affiliation{
Graduate School of Science, Osaka University, Toyonaka Osaka 560-0043, 
Japan\\}
\affiliation{
Graduate School of Engineering Science, Osaka University, 
Toyonaka Osaka 560-8531, Japan
}
\date{\today}
\begin{abstract}
We show that a spontaneous electric current is induced in a nano-scale 
conducting ring just by putting three ferromagnets. 
The current is a direct consequence of the non-commutativity of the spin 
algebra, and is
proportional to the non-coplanarity (chirality) of the 
magnetization vectors. 
The spontaneous current gives a natural 
explanation to the 
chirality-driven anomalous Hall effect.
\end{abstract}
\maketitle

Persistent (permanent) current in metallic rings 
is an equilibrium current which can be 
induced when the time-reversal symmetry is broken \cite{Byers61}. 
 Such a current is expected in the presence of a magnetic flux through 
a normal ring \cite{Buttiker83,Cheung89} and was 
indeed detected experimentally \cite{Levy90,Chandrasekhar91,Mailly93}. 
 The effect is due to a U(1) phase factor attached 
by the flux to the electron wave function. 
 Here we show theoretically that a permanent current is  
induced in a conducting normal ring just by attaching three 
ferromagnets, without magnetic flux through the ring. 
 This surprising effect can be seen  
in nano-scales at low temperatures. 
 The key here is the non-commutativity of the SU(2) spin algebra, which
breaks the time-reversal symmetry, and leads, in the presence 
of electron coherence,  to a permanent electron current.
 Such a system would be utilized in spin polarized transport \cite{Prinz98} or 
in quantum computers as a new kind of logic gates \cite{Feynman82}.

 The electron has spin 1/2 (i.e., has two components),  
and the spin obeys SU(2) algebra. 
 The algebra is represented by three 
$2\times 2$ Pauli matrices 
$\sigma_{i}$ 
($i=x,y,z$) satisfying the commutation relation 
\begin{equation}
[\sigma_i,\sigma_j]
= 2i \epsilon_{ijk}\sigma_{k}, \label{sigcom}
\end{equation}
where $\epsilon_{ijk}$ is the totally antisymmetric tensor with 
$\epsilon_{xyz}=1$. 
 When a conduction electron in a conductor is scattered 
by some magnetic object,
the electron wave function is multiplied by an amplitude 
$A(\nv) = \alpha e^{i\beta\nv\cdot\sigmav}$, 
which is generally spin-dependent and 
is represented by 
a 2$\times$2 matrix in spin space. 
Here $\alpha$ and $\beta$ are complex numbers and $\nv$ is a 
three-component unit vector characterizing the scattering 
object (such as the magnetization direction). 
 We consider in this paper only classical, static scattering objects, 
and assume $\nv$'s are constant vectors.

 Let us consider two successive scattering events 
represented by $A(\nv_1)$ and $A(\nv_2)$ (Fig.1). 
 Due to the non-commutativity of $\sigma_i$, the amplitude depends on the 
order of the scattering event; $A(\nv_1)A(\nv_2) \neq A(\nv_2)A(\nv_1)$ 
in general. 
 Various features in spin transport, which is under intensive pursuit 
recently \cite{Stiles02,Brataas01}, arise from this non-commutativity. 
 It, however, does not affect the charge transport, 
since the charge is given as a sum of the two spin components 
(denoted by tr), and 
$\tr[A(\nv_1)A(\nv_2)] - \tr[A(\nv_2)A(\nv_1)] = 0$. 
 Anomaly in the charge transport arises at the third order.
 We have, by virtue
of Eq.\ref{sigcom} and the relation 
$\tr[\sigma_i \sigma_j] = 2\delta_{ij}$,
\begin{equation}
\tr[A(\nv_1)A(\nv_2)A(\nv_3)] - \tr[A(\nv_3)A(\nv_2)A(\nv_1)]
  = 4 \alpha^3 \sin^3\beta \nv_1 \cdot (\nv_2\times\nv_3) 
  \equiv iC_{123}  , \label{Acom}
\end{equation}
where the cross denotes the vector product, {\it i.e.}, 
$\nv_1\cdot(\nv_2\times\nv_3)=\sum_{ijk}\epsilon_{ijk} n_1^i n_2^j n_3^k$. 
 This relation indicates that in the presence of fixed $\nv_i$'s with
 $ \nv_1 \cdot (\nv_2\times\nv_3) \neq0$, the symmetry under time-reversal 
(more appropriately, reversal of motion) is generally broken in the 
charge transport.  
In fact, relation (\ref{Acom}) indicates that the 
contribution from one path, $x\ra X_1\ra X_2\ra X_3\ra x$ (Fig.1a), 
and its (time-) reversed one, $x\ra X_3\ra X_2\ra X_1\ra x$ (Fig.1b), 
are not equal, and this difference results in a spontaneous electron 
motion in a direction specified by the sign of $C_{123}$, 
namely, a permanent  current. 
 What is essential here is the non-commutativity of the SU(2) algebra. 
 In fact, $C_{123}$ vanishes if all $\nv_i$'s lie in a plane, 
in which case the algebra is reduced to a commutative U(1) algebra. 
 The degree of the symmetry breaking, $\nv_1 \cdot (\nv_2\times\nv_3)$, 
is given by the non-coplanarity, often called spin chirality (Fig.2). 

The spontaneous current above would be realized 
on a small conducting ring with three ferromagnets 
with different magnetization direction, 
$\nv_1$, $\nv_2$ and $\nv_3$. 
 The ferromagnets may be attached to the ring (Fig.3a), or 
embedded in the 
ring (Fig.3b), both being within the reach of present experimental technique. 
 In either case, 
the electron in a ring feels an effective spin polarization 
when it goes through the region ($F_i$) affected by the 
ferromagnets, and the effect will be modeled by the exchange 
(spin-dependent) potential,  
$V(x) = - \Delta \, \nv_i \!\cdot\! \sigmav$ for $x \in F_i$. 
 Here $\Delta$ represents the effective exchange field. 
 The equilibrium charge current in the ring is calculated from 
$ j(x) = \frac{\hbar e}{2m}{\rm Im} (\nabla_{x}-\nabla_{x'}) \tr 
         G(x,x',\tau=0-)|_{x'=x}$, where
$G(x,x',\tau)\equiv-<T c(x,\tau)c^\dagger(x',0)>$ is the thermal Green 
function, $e, m, c$ being the charge, the mass, the 
annihilation operator of electrons, 
respectively.
$G(x,x',\tau)$ is calculated perturbatively from the Dyson equation, $G=g+gVG$, 
where $g$ represents free Green function. 
As is seen from eq. (\ref{Acom}), possible finite current arises at 
the third order in $V$. 
 By summing the contribution of the two paths, 
$x\ra X_1\ra X_2\ra X_3\ra x$ and the reversed one, we have 
$  j(x) = -\frac{\hbar e}{m} B(x) \, {\rm Re} \, C_{123} $. 
 Here $C_{123}$ is defined by eq. (\ref{Acom}) with $\alpha=i\Delta$, $\beta=\pi/2$,  
and 
\begin{equation} 
B(x) =  \prod_{i=1}^{3}\int_{X_i \in F_i} dX_i
   \int \frac{d\omega}{2\pi} 
   f(\omega) \left. \nabla_{X_0} {\rm Im} 
   [ g_{01} g_{12} g_{23} g_{34}] \right|_{X_4 = X_0=x} \label{B}
\end{equation} 
describes the electron propagation through the ring, 
which is common to both paths. 
 In Eq.\ref{B}, 
$f(\omega)$ is the Fermi distribution function and 
$g_{ij} = g^r(X_i-X_j,\omega )$ is the retarded Green's function of 
free electrons.  
 Approximating the transport along the ring as one-dimensional and 
neglecting multiple circulation, we have 
$g^r(x, \omega ) \simeq -i\pi(D/L)e^{ik_F|x|}$, 
where 
$k_F$ is the Fermi wavenumber, 
$D$ the density of states ($\sim 1/\epsilon_F$ ;  
$\epsilon_F = \hbar^2 k_F^2/2m$ being Fermi energy), 
and $L$ the length of the ring perimeter. 
 The final result is given by 
\begin{equation}
  j = -2e \frac{v_F}{L} 
  \cos(k_F L) 
       \left( \frac{J}{\epsilon_F} \right)^3 
       \nv_1 \cdot (\nv_2\times\nv_3), \label{jres}
\end{equation}
at zero temperature. 
 Here $J \equiv \pi W \Delta / L$ with $W$ being the width of the ferromagnets,
and $v_F=\hbar k_F/m$ is the Fermi velocity.

 The current is thus induced by the spin chirality 
$\nv_1\cdot(\nv_2\times\nv_3)$ of the ferromagnets. 
 This quantity reduces to the Pontryagin index (density)
for the case of smoothly varying field $\nv(x)$ \cite{Fradkin91},
which is also interpreted as Berry phase \cite{Berry84} of the 
spin. 
 The effect of spin Berry phase on the electron transport 
has so far been investigated in the limit of strong coupling 
to $\nv (x)$ where the electron spin adiabatically follows 
$\nv (x)$ \cite{Loss90,Ye99}. 
 In contrast, the present result (\ref{jres}) is obtained in the
 opposite limit; 
we have treated the coupling to $\nv$ perturbatively (weak 
coupling) and 
made no assumption of smoothness on $\nv (x)$. 
To see the relation between the two approaches, 
we follow the reasoning used in ref.\cite{Loss90}.
We write the Hamiltonian of our system 
in a general form using a spatially 
varying polarization $\Delta_{\xv}$ as  
$H=\sum_{\xv}\left[\frac{1}{2m}|\nabla 
c_\xv|^2+\Delta_\xv \nv_\xv\cdot (c^\dagger \sigmav c)_\xv\right]
$ and 
move on to a gauge transformed frame, 
$\tilde{c_{\xv}}\equiv U_\xv c_\xv$, with 
$U_\xv=\Mv_{\xv} \cdot\sigmav$. Here
$\Mv_{\xv}=(\sin\frac{\theta}{2}\cos\phi, \sin\frac{\theta}{2}\sin\phi, 
\cos\frac{\theta}{2})$, 
and $(\theta,\phi)$ is the 
polar coordinate of $\nv_\xv$. 
The Hamiltonian is then written as 
$H=\sum_{\xv}\left[\frac{1}{2m}|(\nabla+i\Av)\tilde{c}|^2
+\Delta_{\xv}\tilde{c}^\dagger \sigma_{z} \tilde{c}\right]$, 
where $\Av\equiv -iU^\dagger \nabla U$ is an SU(2) gauge field. 
In the adiabatic limit 
(infinitely large and uniform $\Delta_\xv$) on one hand, only the majority spin 
channel (denoted by $+$) becomes relevant and hence the gauge 
field reduces to a U(1) field, $\av\equiv \Av_{++}$ ($++$-component 
of $\Av$). Thus the new 
electron, $\tilde{c}$, subject to a U(1) magnetic field 
($\nabla\times\av$), exhibits a persistent current.  
In the present weakly coupled case on the other hand, both of 
the two 
spin channels in $\tilde c$ are relevant and so off-diagonal 
components and SU(2) nature of $\Av$ become essential. 
Finite and spatially varying $\Delta_{\xv}$ complicates the problem 
further.
Thus, within the adiabatic scheme, the appearance of persistent 
current in the perturbative regime is not obvious. 
Our result indicates that even in the perturbative  
regime, the electron feels an analog of spin Berry phase and 
the transport is modified.

 The spontaneous current considered here  
is due to the memory during successive 
scattering events, and the electron coherence over the system is essential. 
 The current diminishes as the temperature and/or the ring size 
are increased, as in the conventional persistent current due to 
the external magnetic flux \cite{Cheung89}. 
 The present current is also 
an equilibrium current, which cannot be measured by electrical means. 
 At present it can only be measured by detecting its magnetic moment. 
 The conventional persistent current 
was observed on a single ring of 
gold \cite{Chandrasekhar91} and of GaAs-AlGaAs \cite{Mailly93}. 
 Compared to the conventional one, 
the persistent current proposed in this paper will be smaller 
in magnitude by a factor of $( J / \epsilon_F )^3$. 
 Careful extraction of the magnetic signal due to the 
persistent current from those of ferromagnets 
will also be required in the present case. 
 This may be carried out by fixing one of the magnetizations, say, 
$\nv_1$, perpendicularly to the plane containing the ring, 
keeping $\nv_2$ and $\nv_3$ in the plane. 
 By changing the mutual angle between 
$\nv_2$ and $\nv_3$ in the plane, the spin chirality is controlled without 
affecting the perpendicular component of the magnetic moment. 
 Measurement on many rings on a ferromagnetic network 
will be effective in amplifying the signal. 
(Total signal from $N$ rings grows by a factor $\sim\sqrt{N}$ 
even in the worst case where the sign of the current, 
represented by $\cos k_F L$, is random.) 

The phenomenon predicted here is not restricted to artificial nano-structures, 
but will be present rather generally in metallic frustrated spin systems such 
as pyrochlore ferromagnets and spin glasses, where finite spin chirality is 
often realized \cite{Taguchi01,Kawamura92}.
 The spin chirality was recently pointed out \cite{Ye99}, 
in the adiabatic limit, to be 
the origin of the peculiar anomalous Hall effect observed in experiments 
\cite{Taguchi01}.
 The present chirality-driven persistent current 
affords an intuitive interpretation to it. 
 The circulating current 
starts to drift when the electric field is applied, in the direction 
perpendicular to the electric field (Fig.4), 
just as in the normal Hall effect. 
 With the frequency of the circulating motion, read from Eq.\ref{jres} as 
$ \Omega \simeq 
 \frac{2\pi v_F}{L} \left( \frac{J}{\epsilon_F} \right)^3 
 \nv_1 \cdot ( \nv_2 \times \nv_3 )$, 
we may estimate the Hall conductivity 
by $\sigma_{xy} = \sigma_0 \Omega\tau$ \cite{Kittel86}. 
 Here $\sigma_0$ is the classical (Boltzmann) conductivity, 
$\tau$ is the elastic lifetime, 
and the dirty case $\Omega \tau \ll 1$ is assumed. 
 If the spin chirality is located uniformly on every triangle of size of 
inter-atomic distance (i.e., $\nv_1 \cdot (\nv_2\times\nv_3)= \chi_0$ and 
$L\sim 1/k_F$), we have 
$\sigma_{xy}/\sigma_0 \simeq \chi_0 J^3 \tau/\epsilon_F^2$. 
 This result agrees with the one obtained based on the linear 
response theory \cite{TK02}.

To summarize, we have shown that a spin chirality generally 
accompanies a permanent electric current. 
This effect is a direct consequence of the spin SU(2) algebra. 

\acknowledgements
We thank J. Inoue, H. Akai, H. Fukuyama and T. Ono 
for discussion. 
G.T. thanks the Mitsubishi foundation for financial support.


\begin{figure}[bthp]
%
%
\caption{ A closed path contributing to the 
amplitude of the electron propagation from $x$ to $x$.
 At $X_i$, the electron experiences a scattering represented by an 
SU(2) amplitude, $A(\nv_i)$. 
 The contribution from one 
path (a) and the reversed one (b) 
are different in general due to the non-commutativity of $A(\nv_i)$'s.
\label{Fig1}}

\caption{ Three magnetization vectors with a finite chirality, 
$\nv_1 \cdot (\nv_2 \times \nv_3)$. 
\label{Fig2}}

\caption{Setup for the chirality-driven permanent current.
(a): Three insulating ferromagnets are put on a normal conducting ring. 
(b): Three metallic ferromagnets are embedded in a ring. 
 In both cases, a conduction electron will feel the effective exchange 
field as it goes through the region ($F_i$) affected by the ferromagnets. 
\label{Fig3}}

\caption{ Schematic picture 
of the chirality-induced Hall effect. 
 Circulating permanent currents drift under applied electric field. 
 The crosses denote impurity scattering. 
\label{Fig4}}%

\end{figure}
\end{document}